\documentclass[9pt,twocolumn,twoside]{osajnl}

\journal{ao} % Choose journal (ao, aop, josaa, josab, ol)

\ifthenelse{\boolean{shortarticle}}{\colorlet{color2}{color2b}}{\colorlet{color2}{color2}} % Automatically switches colors for short articles

\usepackage{csquotes}
\usepackage{xspace}
\usepackage{color}
\renewcommand{\deg}{\ensuremath{^{\circ}}\xspace}
\def\fdeg{\hbox{$.\!\!^\circ$}}  %dvg
\def\farcsec{\hbox{$.\!\!^{\prime\prime}$}} %dvg

\title{A fast, wide-field and distortion-free telescope with
 curved detectors for surveys at ultra-low surface brightness}

\author[1,2,*]{Eduard Muslimov}
\author[3,4,5]{David Valls-Gabaud}
\author[1]{G\'erard Lema\^{i}tre}
\author[1]{Emmanuel Hugot}
\author[1]{Wilfred Jahn}
\author[1]{Simona Lombardo}
\author[6]{Xin Wang}
\author[1]{Pascal Vola}
\author[1]{Marc Ferrari}

\affil[1]{Aix Marseille Univ., CNRS, LAM,
 Laboratoire d'Astrophysique
 de Marseille, 38, rue Joliot-Curie, 13388 Marseille, France} \affil[2]{Kazan National Research Technical University named
 after A.N. Tupolev-KAI, 10 Karl Marx, Kazan 420111,
 Russian Federation}
\affil[3]{LERMA, CNRS, PSL, Observatoire de Paris, 61 Avenue
 de l'Observatoire, 75014 Paris, France}
\affil[4]{Institute of Astronomy, Madingley Road,
 Cambridge CB3  0HA, United Kingdom}
\affil[5]{Key Laboratory for Particle Astrophysics,
Institute of High Energy Physics, Chinese Academy of Sciences,
 19B Yuquan Road, Beijing 100049, China}
\affil[6]{Shanghai Institute of Technical Physics,
 500 Yutian Road, Shanghai 200083, China}

\affil[*]{Corresponding author: \url{eduard.muslinov@lam.fr}}

\dates{Compiled \today}

\ociscodes{%
 (080.4228) Nonspherical mirror surfaces;
 (110.0110) Imaging systems;
 (110.6770) Telescopes;
 (040.1520) CCD, charge-coupled devices. }

\doi{\url{http://dx.doi.org/10.1364/ao.XX.XXXXXX}}

\begin{abstract}
We present the design of an all-reflective, bi-folded Schmidt  telescope aimed at surveys of extended astronomical objects with extremely-low surface brightness. The design leads to a high image quality without any diffracting spider,  a large aperture and field of view, and a small central obstruction which barely alters the PSF. As an example, we design a high-quality, 36 cm diameter, fast ($f$/2.5) telescope working in the visible with a large field of view ($1\fdeg6\times2\fdeg6$). The telescope can operate with a curved detector (or with a flat detector with a field flattener) and a set of filters. The entrance mirror is anamorphic and replaces the classical Schmidt entrance corrector plate. We show that this anamorphic primary mirror can be manufactured through stress polishing, avoiding high spatial frequency errors, and tested with a simple interferometer scheme. This prototype is intended to serve as a fast-track scientific and technological pathfinder for the future space-based MESSIER mission.
\end{abstract}

\setboolean{displaycopyright}{false}

\begin{document}

\maketitle
\thispagestyle{fancy}

\ifthenelse{\boolean{shortarticle}}{\ifthenelse{\boolean{singlecolumn}}{\abscontentformatted}{\abscontent}}{}

\section{Introduction}
\label{sec:introduction}

\subsection{Scientific rationale}
The current $\Lambda$CDM paradigm of structure formation has been
well tested on linear scales (\textsl{e.g.}, the fluctuations of the Cosmic Microwave Background), and leads to a very successful theory of
galaxy formation and evolution, although serious problems arise on mildly non-linear scales (\textsl{e.g.}, the putative cosmic web of filaments) and highly non-linear scales (\textsl{e.g.}, number of
satellite galaxies, see \cite{bullock17}). In spite of their widely different scales and densities, they share the common
property of having extremely low surface brightness (LSB), well below the ground-based sky background. While some
instruments are able to uncover new galaxies and LSB features (\textsl{e.g.}, \cite{dmd,vandokkum,trujillo,dragonfly,mihos}),
 a systematic study of the ultra-low surface brightness universe requires large surveys from space. The reasons are obvious: (1) a specially-designed
 instrumentation will maximise the signal-to-noise ratio (SNR) of
their detection, (2) exploiting the much lower sky brightness in space, (3)
 compact PSFs (the scattering produced by atmospheric molecules yields
very extended PSF wings), (4) systematics and accurate flat-fielding can be measured accurately. All these factors are required for a proper photometry of LSB objects.
 This is the basic concept behind MESSIER, a space mission optimised to measure the ultra-low surface brightness universe \cite{messier}. The mission has a number of tight constraints for its design in
 order to achieve its aims of measuring surface brightness levels as low as
35 mag arcsec$^{-2}$ in the optical (350-1000 nm)
and 38 mag arcsec$^{-2}$ in the UV (200 nm). First, the accuracy
of the flat fielding should reach the 0.0025\% level, which is only achievable
through time-delay integration or drift scanning (see, \textsl{e.g.} \cite{gcc}). Second, the optical system cannot include lenses, as they will produce a massive Cherenkov emission by the abundant relativistic
particles. Third, the PSF must be as compact as possible
 and no high spatial frequency errors are acceptable on the optical surfaces. This implies that support structures such
 as spiders cannot be placed in the optical train, while
 the drift-scan mode requires the optical design to be distortion-free at least in one direction.

The goals of the MESSIER mission are achievable with a high-quality, 50 cm diameter,
fast ($f$/2) UV-visible telescope with a huge field of
view ($>2\deg\times 4\deg$). In this paper, we consider a
reduced version of the telescope, intended to serve as a pathfinder for the main mission.
The manufacturing of the pathfinder represents a fast-track project,
aimed at testing and improving a few key technologies required
 for the full-scale project implementation, hence some mitigation
 of the main requirements is allowed. Taking into account  the
 currently available technologies and components, the telescope
 target specifications are defined as follows (see Table~\ref{table1}).
{The resolution requirement is defined by the useable pixel size and approximately corresponds to 2\farcsec5 per pixel. Using the rms spot radius is a simplified but convenient way to estimate the optical quality during design and optimisation,
while controlling the spot size makes the
PSF core to match the size of the pixel. To achieve the most compact shape of the PSF we should minimize the obscuration and exclude spiders at the design stage.}

\begin{table}[htbp]
\centering
\caption{\textbf Main parameters for the MESSIER pathfinder telescope.}
\begin{tabular}{p{1.3cm}p{1.0cm}p{2.5cm}p{2.5cm}}
\hline
Parameter & Symbol & Value       & Value \\
          &        & (pathfinder) & (space telescope)\\
\hline
Field of view    &   FoV       & $1\fdeg6\times 2\fdeg6$    & $2\deg\times 4\deg$ \\
$f$-ratio        & $\Omega$    & 2.5                       & 2.0 \\
Primary diameter &   M1        & 356 mm                    & 500 mm \\
Distortion       &$\epsilon$   & $<$0.5\% & $<$0.5\% \\
                 &        & (in one direction) & (in one direction)\\
Spot rms radius  &$\rho$       & $<$12$\mu$m               & 5--15$\mu$m\\
\hline
\end{tabular}
  \label{table1}
\end{table}

In this context, many different options for the optical design of such survey telescopes have been considered (see \textsl{e.g.}
\cite{buffington,terebizh11,terebizh16}). In particular, off-axis optical designs such as the Three Mirror Anastigmat (TMA) with aspherical \cite{cabanac,chang06,chang13}, or freeform mirrors surfaces \cite{challita,agocs} yield a very high performance over an extended field of view (FoV).
 These designs were analysed in detail and compared with other schemes such as the Schmidt and SEAL telescopes
 \cite{hugot:messier,lemaitre14}, and led to the conclusion that the modified reflective Schmidt design fulfils all the requirements
 in terms of optical performance and, in addition, is made up of easily  manufacturable optical components. Further, the simple
 optical scheme eases the telescope assembling and alignment.

In this paper we therefore study the properties, limitations and manufacturability of this reflective Schmidt optical design.
We first present the basic bi-folded Schmidt design, and consider
and quantify the advantages and shortcomings inherent to this type
 of telescopes.  We then describe a design to reach to the
 specifications given above for a MESSIER pathfinder telescope,
 and  study the anamorphic aspherical mirror used in this scheme,
 providing details on its manufacturability by
 active optics methods and its optical quality testing.

\section{BI-FOLDED SCHMIDT DESIGN}
\subsection{The design concept}
\begin{figure}[htbp]
\centering
\fbox{\includegraphics[width=\linewidth]{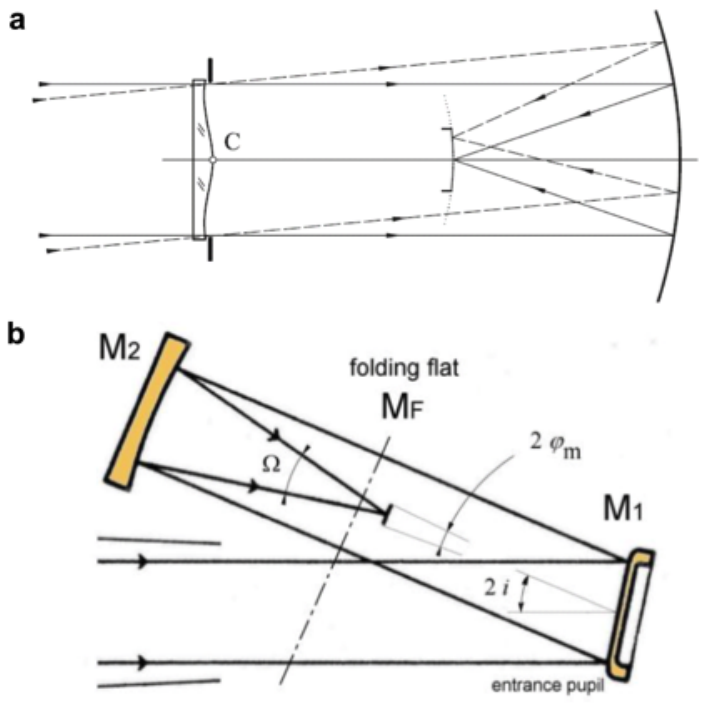}}
\caption{
The main optical schemes of Schmidt telescopes.
\textbf{a} Refractive design: the basic telescope layout with a refractive corrector plate and a curved focal plane.
\textbf{b} Bi-reflective Schmidt configuration with the definition of its basic parameters.}
\label{figure1}
\end{figure}

The basic principle of the wide field telescope invented by B.~ Schmidt is
 a single concave spherical mirror, used with an entrance pupil
 stop at its centre of curvature \cite{schmidt} {(though a similar principle was considered by a few authors before him)}. This combination
 is corrected of each aberration, over the FoV, except for
 spherical aberration. To compensate, an entrance refractive corrector plate is placed at the entrance pupil that provides
 corrected images over all the field of view
(Fig.~\ref{figure1}a).  By balancing the slopes of the aspherical
plate, \textsl{i.e.}, the 1st-order derivative of the aspherical
plate, the chromatic variation of spherical aberration
(sphero-chromatism) remains the dominating aberration residual.
Another basic feature of the Schmidt design is the advantage of
a curved focal surface on which the celestial sphere is
\textsl{naturally} projected, resulting in a projection that is
entirely free
from distortion aberration, and making it ideally-suited
for wide-field telescopes aimed at sky surveys.
Similarly, the proposed all-reflective design shown in
Fig.~\ref{figure1}b uses an off-axis reflective aspheric
corrector, so that sphero-chromatic blur aberrations disappear.
In his analyses of reflective Schmidt designs for telescopes and
 spectrographs, Lema\^{i}tre (1979,\cite{lemaitre79}) showed that
 in all-reflective designs the best angular resolution over the field of view is achieved by balancing the meridian curvatures,
 \textsl{i.e.} the 2nd-order derivatives, of the aspherical mirror
 or diffraction grating \cite{lemaitre79,lemaitre09}.
 If, as we do in this paper, one assumes an incident beam with a
circular cross section at the entrance pupil aspheric mirror
(primary mirror M1) of clear aperture $2R_m$, the curvature balance
rule states that any parallel ray within a cylindrical tube
of aperture $2R_0$ will reach an elliptical null-power zone in M1
such that the relationship between the $R_0$ and $R_m$ radii
is given by the simple ratio
\begin{equation}
\frac{R_0}{R_m} \; = \; \sqrt{\frac{3}{2}} \; \approx \; 1.224 \; .
\label{eq1}
\end{equation}
This leads to an elliptical null-power zone outside the clear aperture. Incidentally, this result is very different from the
case of a refractor plate where the null-power zone is such
that $R_0/R_m = \sqrt{3}/2 \approx 0.866$, \textsl{i.e., inside}
the clear aperture of the beam.  Hence the optical surface of the M1 mirror is
 generated by homothetic ellipses having principal axis lengths in
the ratio $\cos i$. The principal axis lengths of the elliptical
clear aperture entrance pupil M1 are $2R_0$ and $2R_0/\cos i$.
Such reflective Schmidt design geometries have been adopted for broad-band faint object spectrographs with aspherised gratings
\cite{lemaitre09} and also for the giant Chinese segmented
telescope LAMOST \cite{lamost:optics,lamost:main}. However, in these
cases, the focal surface in the middle of the beam creates a
central obscuration and additional diffraction spikes are caused
by the holding spider.c
Therefore, in our proposed design, both the spiders and the corresponding diffractive deformation of the PSF are eliminated
by adding a folding flat mirror. This also allows a much
better access to the focal surface, which is placed outside
the optical train (see Fig.~\ref{figure1}b).

In the case of a purely reflective Schmidt, except for particular cases such as aspherised grating cameras working in reflected
normal-diffraction \cite{lemaitre09}, a rotational symmetric
corrector does not reach the optimal resolution. Indeed, a 5th-order astigmatism aberration ($\textrm{Astm}_5$) residual remains
all over the field of view and thus is not corrected at the centre.
 A much higher angular resolution can be achieved through a
 non-axisymmetric corrector where the M1 surface has homothetic elliptic level lines. When the extreme curvatures along meridian sections of the aspheric are balanced (Eq.~\ref{eq1})
 --algebraically opposite at centre and edge--  the resolving
 power $d_{NC}$ is given by \cite{lemaitre79,lemaitre09}
\begin{equation}
d_{NC} \; = \; \frac{3}{256 \, \Omega^3} \, \left( \varphi_m^2
+\frac{3}{2}\, i \, \varphi_m\right) \; ,
\label{eq2}
\end{equation}
where $\Omega$ is the focal ratio, $i$ the folding angle
and $\varphi_m$ the maximal semi-field angle (Fig.~\ref{figure1}b).
 Hence the resolution degrades not only with increasing
 the aperture and/or the field of view, but also with the folding
 angle. Note that all the angles and distances in the optical
 scheme should
 provide gaps between components accounting for their real thickness, mechanical holders and other parts.
 For a convenient angular resolution, Eq.~\ref{eq2} implies
 exploring  designs with $f$-ratios $\Omega=2-2.5$, field angles in perpendicular direction to the telescope symmetry plane $2\varphi_m=2\fdeg5-5\deg$, and folding angles $i = 10\deg-13\deg$. In the next section we consider a number of possible designs and investigate the factors limiting their performance.

\subsection{Exploration of parameter space}

\begin{figure}[htbp]
\centering
\fbox{\includegraphics[width=\linewidth]{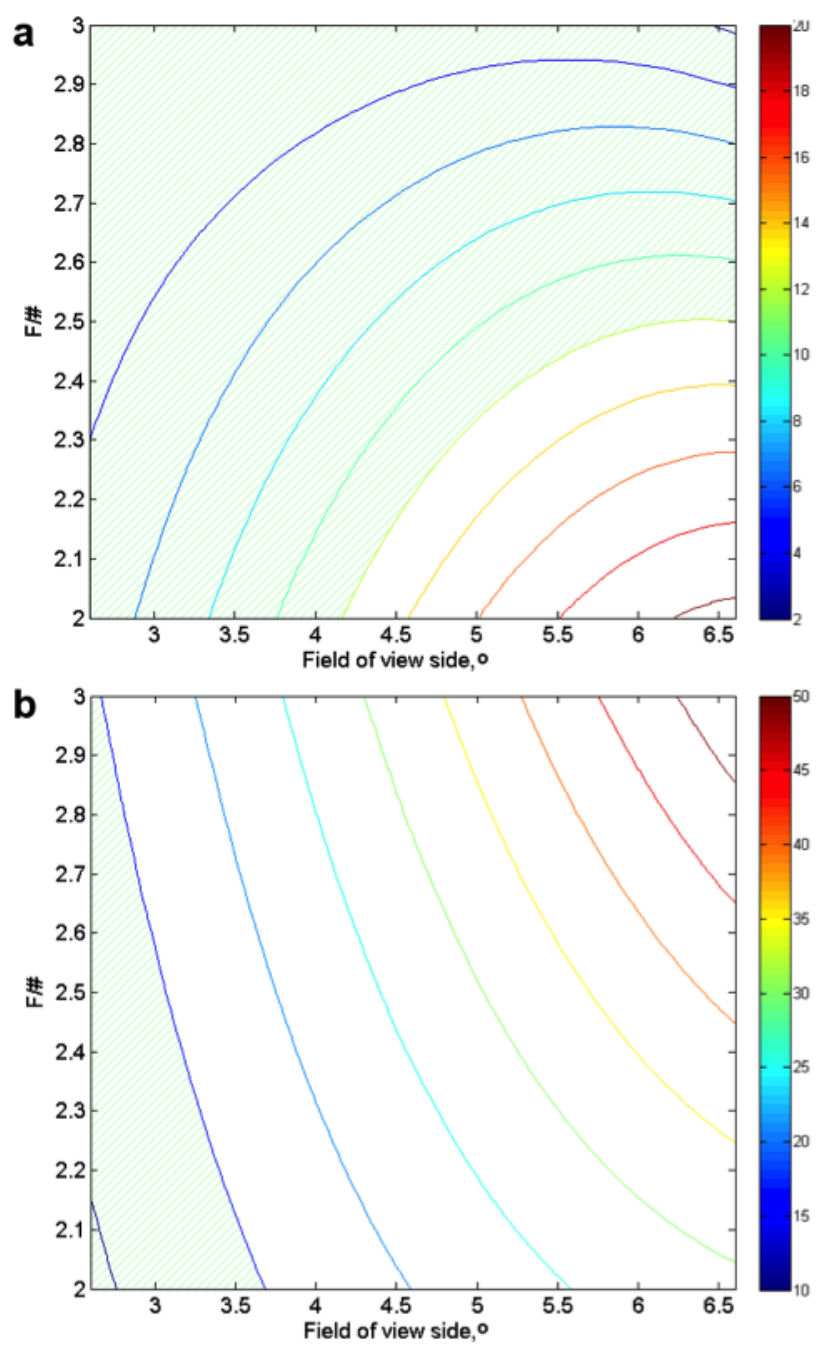}}
\caption{Dependence of the basic parameters of a purely reflective Schmidt telescope as a function of $f$-ratio and field of view (in degrees): \textbf{a} RMS spot radius (in microns) at the edge of the FoV, \textbf{b} Central obscuration (in percentage).}
\label{figure2}
\end{figure}

To quantify the above-mentioned limiting factors a few similar
bi-folded Schmidt schemes were designed, all of them with
 square fields of view. Each time
the general configuration was changed, the angle of rotation of
the 1st and 2nd mirrors were modified as well to avoid ray obscuration by the back side of any of the mirrors.
When necessary, the distances between the mirrors were also adjusted. Each system was numerically optimised to provide high
 optical quality on a curved surface. In total, some 15 different configurations varying the FoV and $f$-ratio were considered.
We found that the optical quality at the field centre is relatively stable and never overcomes the limit of 12$\mu$m set by the requirements (Table~\ref{table1}), while the spot radius at
the edge of the field of view breaks this limit at some point. The distortion
never exceeds the limit if it is measured along one axis of the
square field of view.
% The central obscuration grows significantly with
% change of the general parameters.
 Other design parameters such as the asphericity of the primary mirror and the size of the telescope detector curvature do not change significantly and never reach critical values.
The resulting numerical data were fitted with simple polynomial approximations to derive the dependencies of the resolution and central obscuration on the field of view and $f$-ratio values
(see Fig.~\ref{figure2}a and b, respectively). The hatched areas  correspond to spot RMS radii smaller than 12$\mu$m (Fig.~\ref{figure2}a) and central obscurations smaller than 15\% (Fig.~\ref{figure2}b). This critical value is rather arbitrary as
there is no obscuration requirement which follows directly
from the scientific and engineering goals.
Figure~\ref{figure2} shows that the central obscuration grows rapidly when the FoV and aperture are increased, while the resolution remains acceptable over a large area in parameters space.
The central obscuration is therefore the main limiting factor for this type of optical designs, and it can be loosened to some
extent using decentered and/or elliptical openings in the secondary mirror, but it cannot be completely eliminated.
Another undesired effect observed when the FoV and $f$-ratio are increased is the decrease of the distance from the
back side of the secondary mirror to the detector. This effect is not taken into account yet, as some critical values and limits --not yet defined-- strongly depend upon the type of detector and on
the conditions of its use.
The behaviour of these quantities in parameter space were
used when designing the MESSIER pathfinder optical scheme, and
 they can also be very useful for any future developments
 exploiting the same concept.

\section{TELESCOPE DESIGN}
\subsection{Optical scheme}
In contrast with the full-scale space-based MESSIER telescope,
the ground-based pathfinder must
be delivered in a relatively short time. Under these conditions,
two versions of the optical scheme can be considered:

\textbf{Configuration I}: The focal surface is covered with a curved detector, and no auxiliary optics is used. This version
corresponds to the instrument for the future MESSIER mission,
 but has higher technological risks.

\textbf{Configuration II}: The telescope uses a
conventional ordinary flat detector and a refractive field flattener. This version has the highest tecnological readiness level, but cannot used for space applications. In this case a
filter wheel must be added to enhance the contrast and reduce the field-flattener chromatism.

\begin{figure}[htbp]
\centering
\fbox{\includegraphics[width=\linewidth]{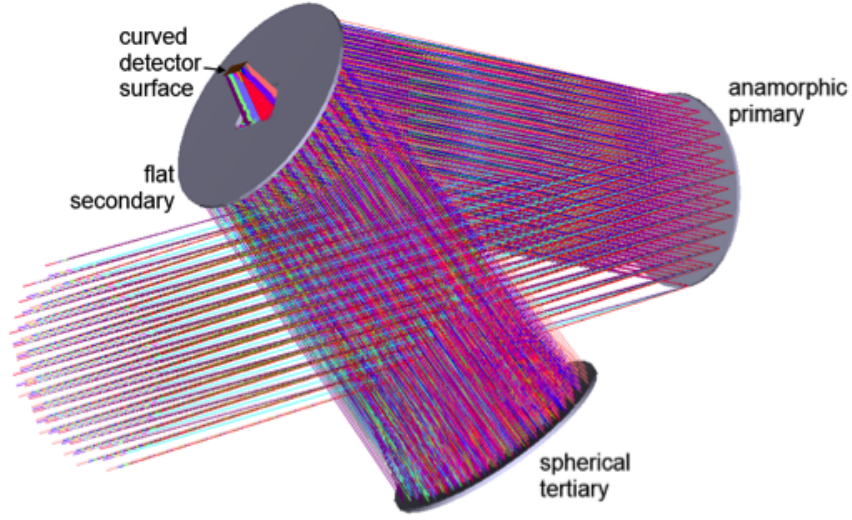}}
\caption{Optical scheme of the bi-folded reflective Schmidt telescope. }
\label{figure3}
\end{figure}

Figure~\ref{figure3} shows the general view of the telescope
 scheme in Configuration I. The folding angle in this case
 is $i = 11\deg$, the rest of the main design parameters correspond to those in Table~\ref{table1}.

The geometry of Configuration II is almost the same but a
symmetrical bi-convex field-flattener lens ($R$=635.75 mm,
$d$=4 mm in fused silica) and a filter wheel are introduced
directly in front of the detector. We adopt the SDSS filter
set \cite{sdss} in the $u, g, r, i$ and $z$ passbands.
A central opening in the secondary mirror is implemented in the form of a decentred trapezoid to decrease the obscuration losses.
This configuration change requires a shift of both the spherical mirror and the detector unit. The change from  Configuration I
to Configuration II implies that the secondary-to-tertiary
distance should be decreased by 31.78 mm, while the distance from the tertiary to the detector increases by 2.96 mm.
These modifications are relatively small and can easily be
implemented in the same optomechanical design.

\subsection{Image quality}
\begin{figure}[htbp]
\centering
\fbox{\includegraphics[width=\linewidth]{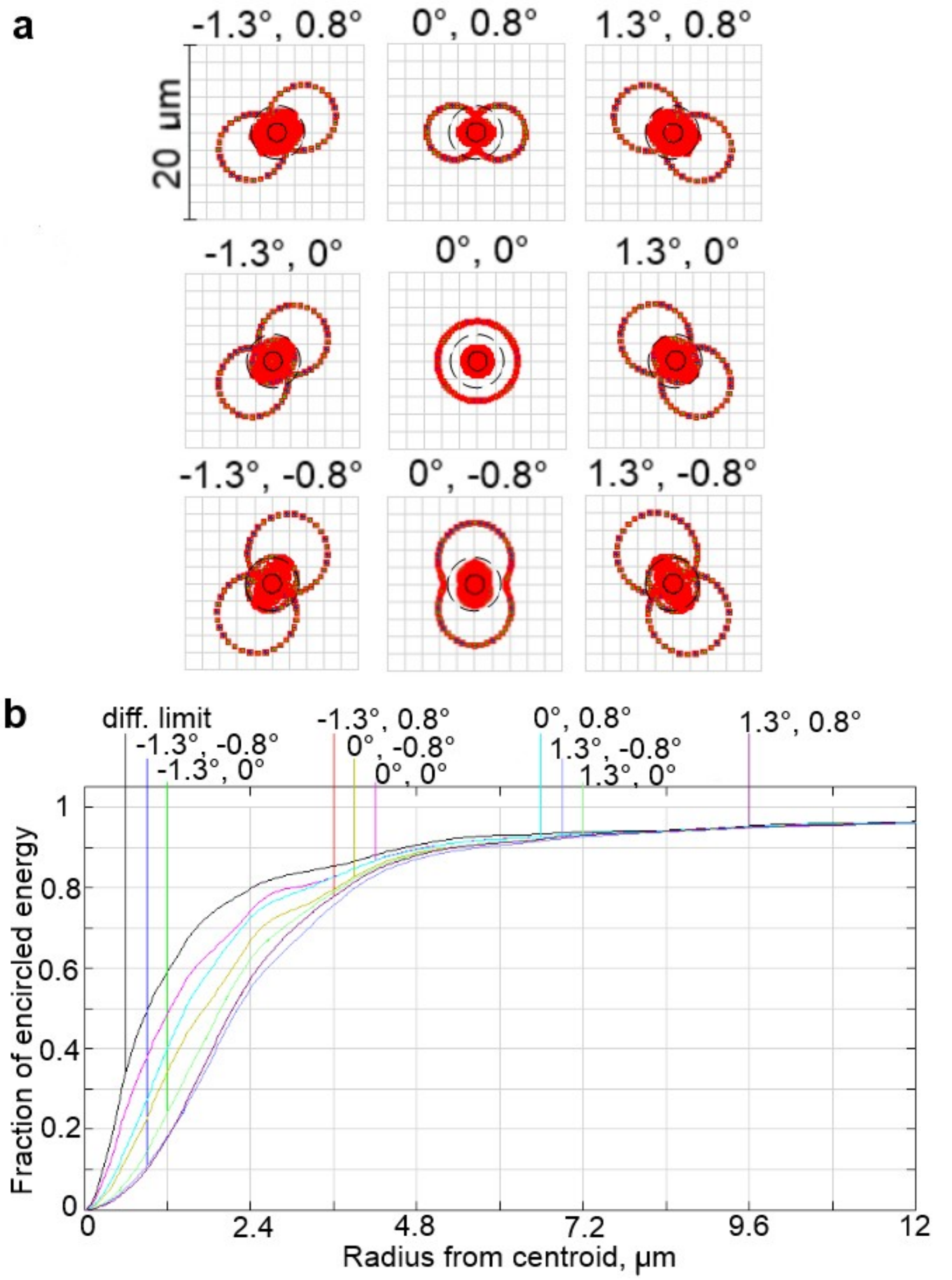}}
\caption{Telescope image quality in Configuration I:
\textbf{a} Spot diagrams {(the solid circle corresponds to the Airy disk at $\lambda$=350 nm, \textit{r}=1.1  $\mu$m ), the dashed circle is that at $\lambda$=1000 nm, \textit{r}=3.1  $\mu$m)}, \textbf{b} Diffraction encircled energy. }
\label{figure4}
\end{figure}

The image quality of the telescope is measured  through
 spot size diagrams. The
  diagrams for Configuration I  (Fig.~\ref{figure4}a) show
  that the image quality is close to the diffraction limit.
 In Figure~\ref{figure4}b, the energy concentration plots  demonstrate that the contribution of the residual aberrations
 observed on the spot diagrams is negligible.
The spots radii for both configurations are given
in Table~\ref{table2}. {Note the  high uniformity of the image quality across the FoV.}

\begin{table}[htbp]
\centering
\caption{\textbf Image quality of the optical design with filters.}
\begin{tabular}{ccc}
\hline
Waveband limits  & Spot rms radius & Spot rms radius \\%
 $\, $ [nm]       & at FoV center [$\mu$m] & at FoV edge [$\mu$m]\\%
\hline
\multicolumn{3}{c}{Configuration I} \\
\hline
350 -- 1000    &   2.1       & 3.1 \\
\hline
\multicolumn{3}{c}{Configuration II} \\
\hline
350 -- 410   &  1.8  & 3.7  \\
400 -- 560   &  2.3  & 3.8   \\
550 -- 700   &  1.7  & 3.4  \\
680 -- 860   &  1.7  & 3.3 \\
860 -- 980   &  1.7  & 3.3 \\
\hline
\end{tabular}
  \label{table2}
\end{table}

The image quality in Configuration II remains almost unchanged and matches the requirements.

\begin{figure}[htbp]
\centering
\fbox{\includegraphics[width=\linewidth]{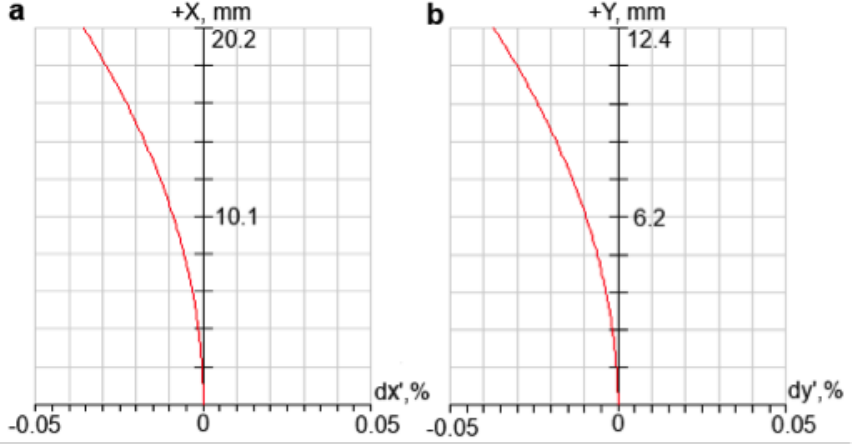}}
\caption{Telescope distortion: \textbf{a} along the $X$ axis,
and \textbf{b} along the $Y$ axis.}
\label{figure5}
\end{figure}

\begin{figure*}[htbp]
\centering
\fbox{\includegraphics[width=\textwidth]{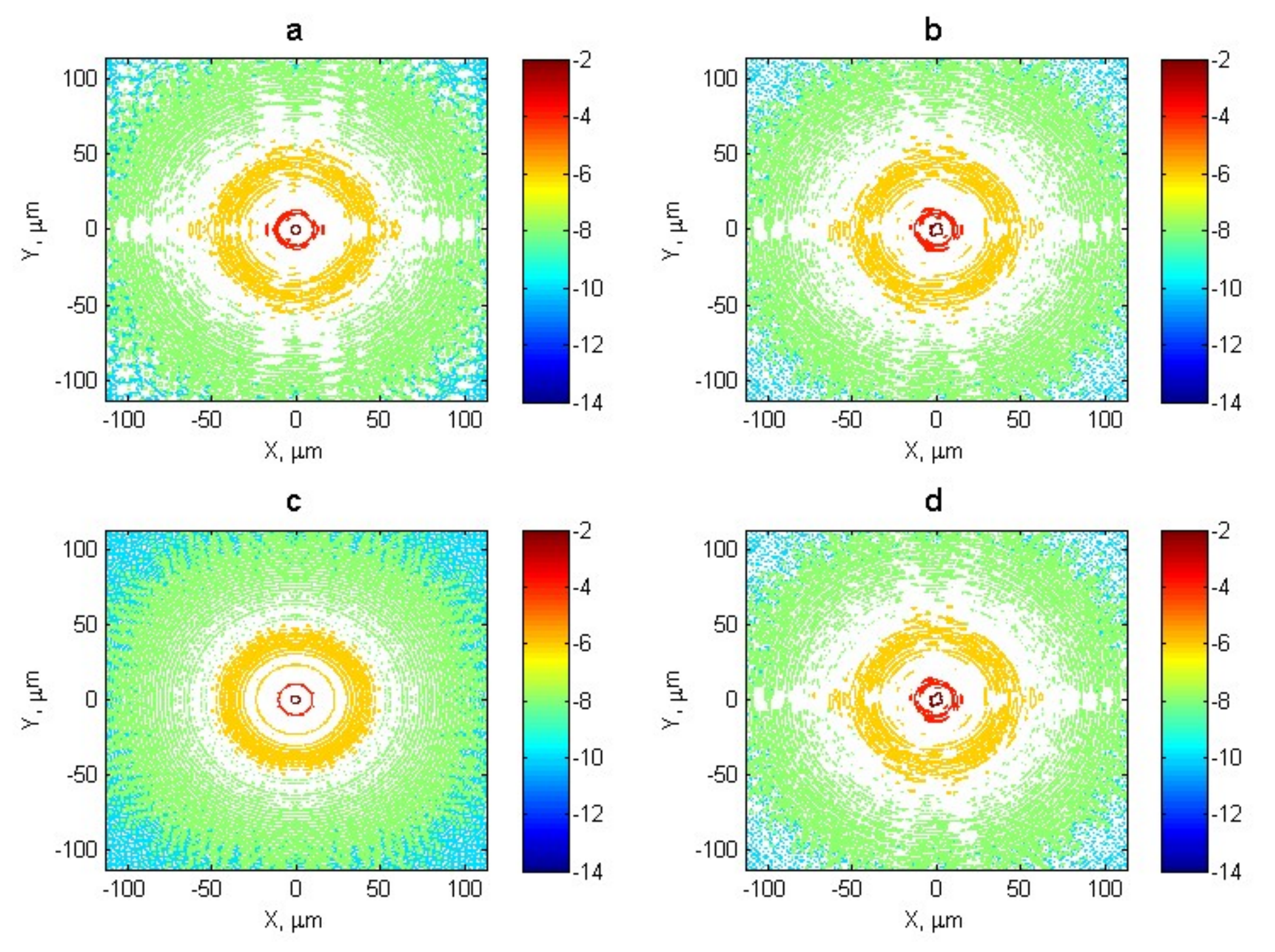}}
\caption{Telescope PSFs in Configuration I: \textbf{a} at the centre (0\deg, 0\deg) of
the FoV with obscuration,
\textbf{b} at a corner (1\fdeg3, 0\fdeg8) of the FoV with obscuration,
\textbf{c} FoV centre without obscuration, \textbf{d} FoV corner without obscuration.}
\label{figure6}
\end{figure*}

Finally, one of the key requirements for the instrument is the absence of distortion. We illustrate the implementation of this
requirement in Fig.~\ref{figure5}.  The correction of distortion over an extended FoV is inherent to Schmidt-type designs, so
the distortion along both axes is insignificant and well
below the requirements.
{We note here that the current design is relatively sensitive to the components misalignments and manufacturing errors. It is an unavoidable property of off-axial optical systems using aspheres. However, preliminary tolerances estimations performed by  standard Monte-Carlo techniques show that all the values are technologically realistic. For instance, the typical limit for mirrors longitudinal shift of the mirrors is 0.1 mm, and for the detector it ranges from 0.02 to 0.1 mm depending upon the assumptions made on the focusing mechanism. The typical limits to the decentering of the components  are 0.05-0.1 mm. The corresponding values for the tilts are 0\fdeg003--0\fdeg1 with the highest sensitivity for the spherical mirror. The tolerances on surface shapes depend strongly on the initial assumptions and calculation techniques. Approximate limits  are 500, 100 and 50 $\mu$m for the primary, secondary and tertiary mirrors, respectively.
It is premature at this stage to provide here a more detailed tolerance analysis, which will
be carried out in a forthcoming study.}

In addition to the usual ordinary image quality estimates,  the
shape of the PSF is an essential factor, and particularly its wings,
for the proper photometry of ultra-low surface brightness objects.
We sampled two points in the FoV: the center at (0\deg, 0\deg) and one corner at (1\fdeg3, 0\deg8). {The Huygens formalism is used to compute the PSF, so the results are normalized to the diffraction limit.    Fig.~\ref{figure6} shows the results, converted to logarithmic scales in both cases (with and without central obscuration taken into account in the computations). One can see from the diagrams that the perturbations on the PSF wings are very small.
We also note that the central obscuration and deviation from the axial symmetry affect the PSFs, but their influence is insignificant outside of the core.}

\begin{figure}[htbp]
\centering
\fbox{\includegraphics[width=\linewidth]{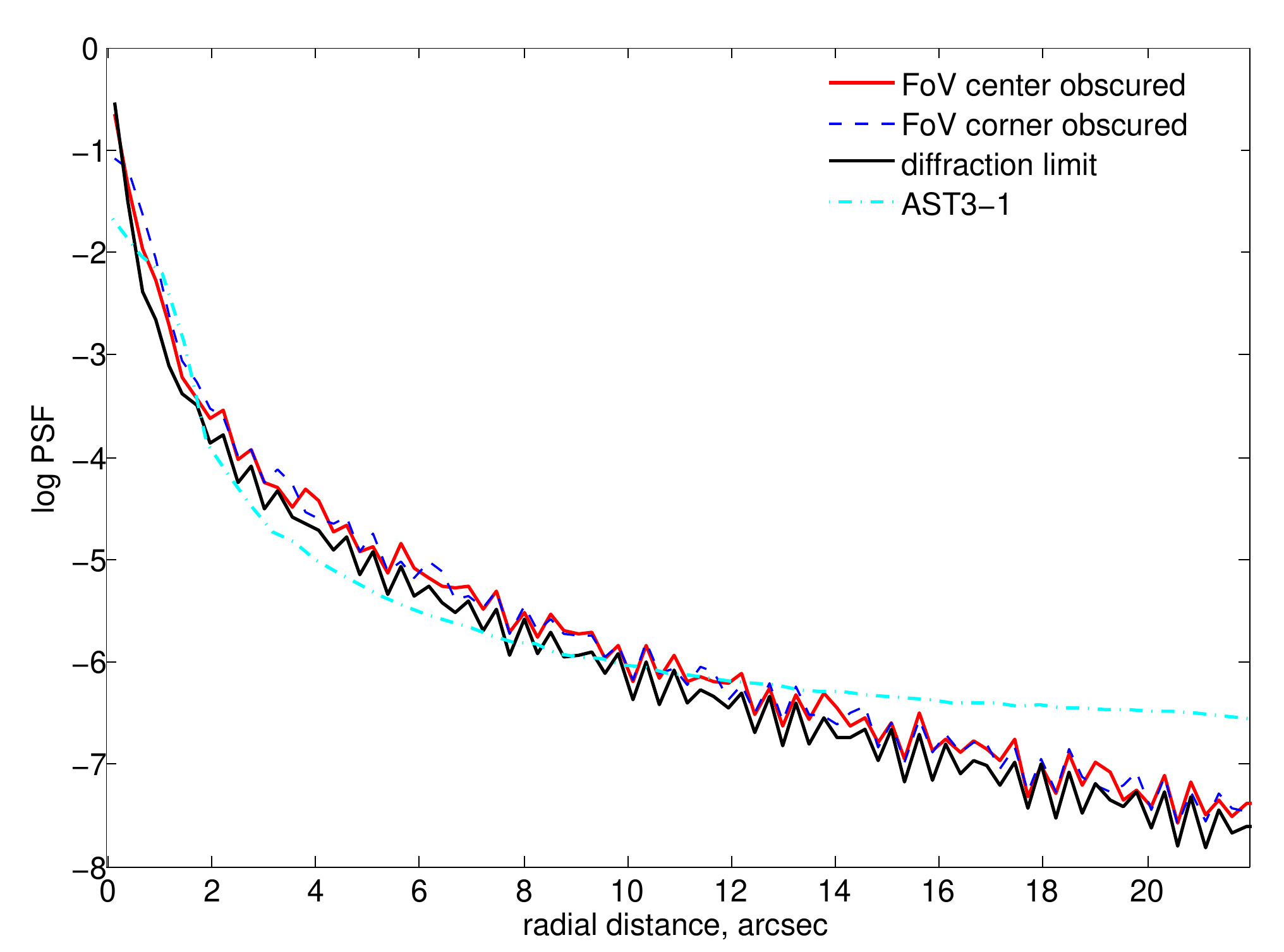}}
\caption{ {Circularised PSF of the MESSIER pathfinder telescope in comparison with (1) the
diffraction limit, and (2) the PSF for the Antarctic Survey Telescope design \cite{yuan}.}}
\label{figure7}
\end{figure}

Figure~\ref{figure7} shows that, after averaging over circular annuli, the PSF for our design of the bi-folded reflective Schmidt
 telescope approaches the diffraction limit {(the diffraction limit curve is re-sampled to facilitate comparison), and at
 this scale the effects of aberrations and central obscuration are almost negligible.
 Compared with the optical scheme of, for example, the Antarctic Survey Telescope  --an axial Schmidt-type design with refractive correctors \cite{yuan}--, the PSF of our
 design is about one order of magnitude better at 22 arcsec, and
at larger angular distances the difference is even more important. In addition, it has also
a larger Strehl ratio.
Therefore our design leads to a notable improvement for
 observations of extremely faint extended objects in
comparison with the existing telescopes of the same class.}

\subsection{Notes on the primary mirror feasibility}
The most complex component in the telescope design is the primary mirror. As mentioned above, when the mirror inclination angle exceeds a certain value it becomes impossible to use an axisymmetric shape of the surface. Its surface should be anamorphic. There are different ways to describe such a surface, and in practice
it is more efficient to use Zernike polynomials when optimising the optical design. However, at the manufacturing and optical testing
phases, it is more convenient to use an anamorphic asphere
equation (a standard option in the ZEMAX optical design software)
\begin{eqnarray}
Z \, &=& \, \frac{C_x \,x^2 \, + \, C_y\,y^2}{
1+\sqrt{1-(1+K_x)C_x^2 x^2 \; - \, (1+K_y)C_y^2 y^2}} \, \\\nonumber
 &+& \,
AR \left[ (1-AP)x^2 \,+\, (1-AP)y^2\right]^2
\label{eq3}
\end{eqnarray}
where ($y, z$) is the symmetry plane of the telescope.
Assuming $K_x=K_y=-1$, the denominator can be excluded, and
in this case the primary mirror shape for the design under consideration is described by just four coefficients:
$C_x=-3.622 \,10^{-6}, C_y=-3.490\,10^{-6}, AR=2.203\,10^{-11}, AP=-0.01854$.

\begin{figure}[htbp]
\centering
\fbox{\includegraphics[width=\linewidth]{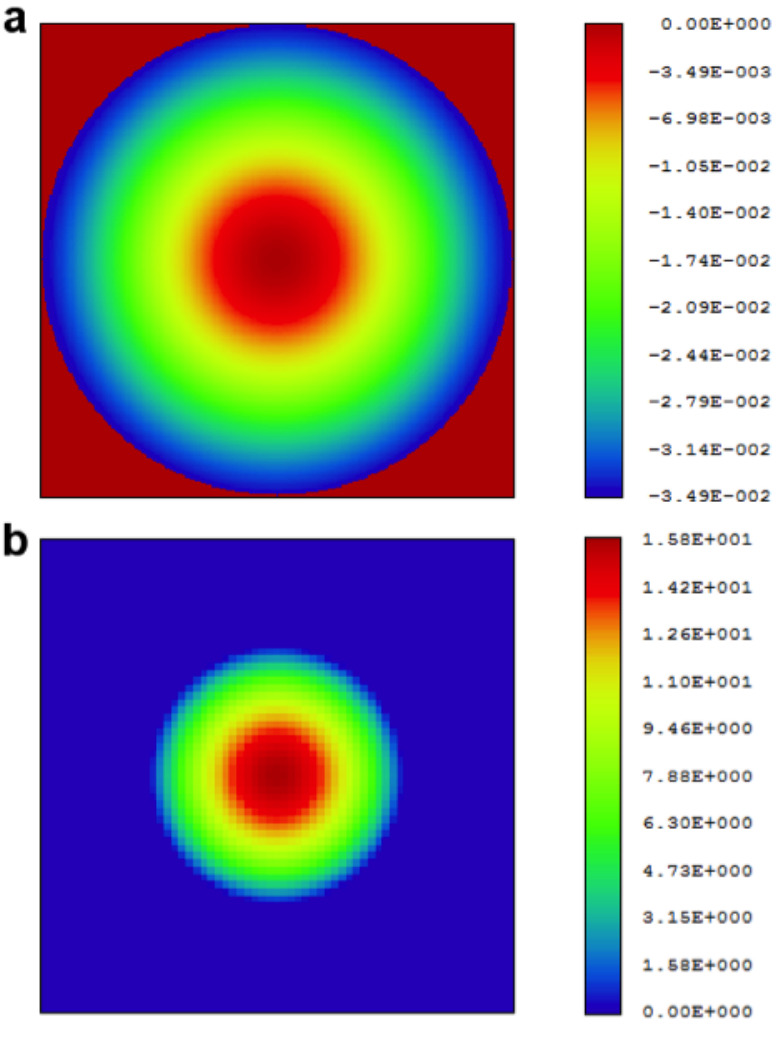}}
\caption{Surface shape of the freeform primary mirror. \textbf{a} Surface sag (in millimeters), \textbf{b} asphericity (in microns).}
\label{figure8}
\end{figure}

It is a \textsl{flat asphere} with a RMS deviation of only 11.9 $\mu$m and a maximum deviation of 19.7 $\mu$m in comparison with the
best-fit sphere (BFS) of radius $2.56\, 10^7$mm. Figure~\ref{figure8}a and \ref{figure8}b show the surface
sag and its deviation from the BFS, respectively.

The mirror is of the freeform class \cite{forbes,hugot:fame}, as
its surface has no axial symmetry and cannot be simply described by conicoids extensions or off-axis conicoids.
This surface can easily be manufactured through active optics methods \cite{lemaitre80,lemaitre09}. We assume that the mirror is the surface of twin vase-form elliptic blanks, made of Zerodur
\cite{pepi14}, Young modulus 90.2 GPa and Poisson ratio 0.243.
These vase forms are glued by a DP490 epoxy resin of 100 $\mu$m, Young modulus 659 MPa and Poisson ratio 0.380. In this particular geometry, called \textsl{closed vase} or \textsl{closed biplate} form,  the radial thickness of assembled outer rings ($t_r$=18 mm with homothetic ratio variation) provides a semi-built-in effect in order to achieve the required aspheric profile over the whole optical clear aperture of the inner constant thickness plate. The outer rings have
an axial thickness of 38 mm and their inner surfaces are
built-in to the plate at the elliptical contour 356$\times$362.7 mm which defines the clear aperture. A round inner corner $R$=5 mm avoids stress concentration at the link.

Next we analyse how to generate the freeform surface with
both basic analytic calculations and by finite-element analysis (FEA).
The FEA model takes into account the bi-symmetry of the mirror:
only a quarter of the mirror is meshed. A fixed node on the central
axis and radially sliding plane degrees of freedom define the
kinetic boundary conditions. When a constant pressure is applied
to the close vase form, the inner plate surface is bent accordingly,
following a bi-quadratic law, \textsl{i.e.} co-adding the curvature
 and 4th degree modes.  The correct shape is achieved
after an elastic relaxation phase which takes place once
the load is removed after polishing.

Additional optimisations using the elasticity code and the optical design code showed that the entire surface of the inner plate
can be used as the clear aperture entrance pupil of the telescope.
 In doing so, the ratio between the null power zone $R_0$ radius
 and the clear aperture $R_m$ radius (\textsl{i.e.} $R_0/R_m = 1.224$ in Eq.\ref{eq1}), is slightly modified and becomes
\begin{equation}
\frac{R_0}{R_m} \, = \,1.118 \; .               \end{equation}
In spite of this decrease, the angular resolution remains
 close to the original value, as the resolution increases
 very slowly for smaller values of this ratio
\cite{lemaitre79}. The entire surface of inner plate can
then be used as clear aperture.
The FEA analysis further shows that a
 constant pressure
$q$=0.687$\times$105 Pa must be applied in the close vase form,
 with inner plates of 18 mm constant thickness during the stress figuring and polishing.

\begin{figure}[ht]
\centering
\fbox{\includegraphics[width=\linewidth]{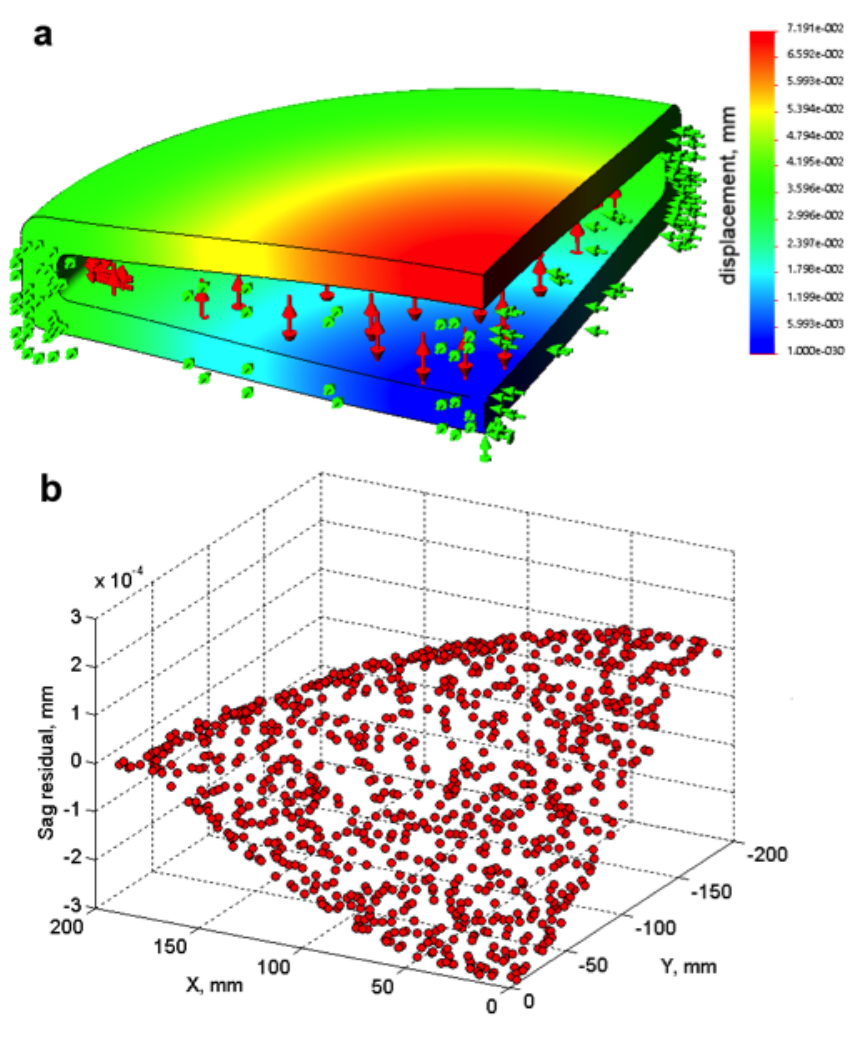}}
\caption{Analysis of the freeform surface as shaped through active optics methods: \textbf{a} finite-element model of the deformed mirror, \textbf{b} residual of the surface sag. }
\label{figure9}
\end{figure}

The general view of the FEA model (Fig.~\ref{figure9}a)
illustrates the close vase form required to achieve the theoretical shape
(defined by Eq.~\ref{eq3}) where the maximum flexural sag from
the center to the elliptical clear aperture is
34.8 $\mu$m with the above parameters.
The surface to be polished reaches the outer contour at 376$\times$383.3 mm,  including a part of the outer elliptical ring
ending with a round corner of radius $R$=8mm. Fig.~\ref{figure9}b shows
the difference, over the entire surface, between  the
flexural sag predicted by the FEA and the one predicted by the
simple theoretical analysis. By removing the ring zone that
does not belong to clear aperture, residual errors become
 negligible. Further details on elasticity analysis  for these
 configurations can be found in \cite{lemaitre80,hugot08,lemaitre:fireball}.

The RMS value of the sag residual shown on Fig.~\ref{figure9}a is 0.13 $\mu$m, while its PTV value is 0.4 $\mu$m. These values are computed for the entire surface, including the outer annulus. For the clear aperture the deviation is even smaller.

\subsection{Notes on the optical testing of the primary mirror}
\begin{figure}[htbp]
\centering
\fbox{\includegraphics[width=\linewidth]{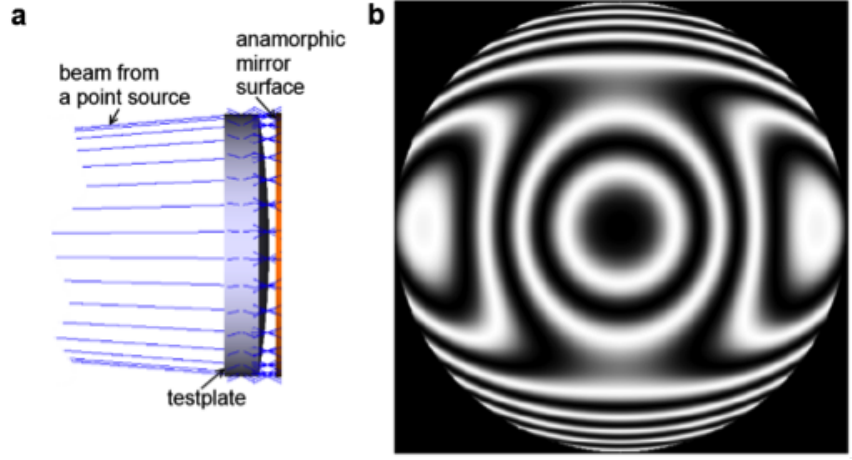}}
\caption{Optical testing of the freeform primary: \textbf{a}  measurement mounting scheme, \textbf{b} simulated interferogram at 632 nm. }
\label{figure10}
\end{figure}

We now turn to the optical testing of the freeform mirror. The precise measurement of a large freeform mirror, with a complex anamorphic shape and almost zero optical power,
could be difficult. Several options for the testing scheme
included the use of lens compensators \cite{malacara}, computer-generated holograms \cite{holo}  and their combinations,
but all of them imply the use of large-format customised optical components. This would not only increase the cost of the
mirror, but also would require more time and expertise in performing the optical testing.
We adopted the alternative of using a single optical element, an existing lab testplate, for the testing scheme.
 The scheme (see Fig.~\ref{figure10}a) operates in autocollimation, when a laser beam emitted by a point source is shaped by the testplate and reflected back by the mirror. The beam path backwards is almost identical to that in the forward direction. Some residual aberrations remain, due to the surface asphericity. However, the simulated interferogram indicates that these residuals correspond to only 7 interference fringes (see Fig.~\ref{figure10}b). Such an interference pattern can be easily measured and then compared with the calculated one to control the surface shape with a high precision. Among other features, the interferogram clearly demonstrates anamorphic shape of the surface.

\section{CURVED DETECTOR OPTIONS}
\label{section4}

The image quality analysis shows that the best performance of
this design is achieved using a curved detector. The
 Configuration I described above
requires a curvature radius of 887.1 mm and an active area of 40$\times$25 mm, which can be achieved within current technologies.
Indeed, over the past few years, there have been several
developments in curved detectors carried out by
Sarnoff \cite{swain}, Stanford \cite{rim,dinyari},
University of Arizona \cite{lesser} and JPL \cite{nikzad}.
The different techniques proposed for the bending of CCD and CMOS
have yielded prototypes with a very low performance loss in terms of dark current, noise, and a filling factor above 80\%.
In astronomy, R\&D activities have been undertaken by ESO for the development of large-format, highly-curved detectors for the
VIS range
\cite{iwert}. The manufacturing process has been fully developed
and has delivered a working component suitable for astronomical applications. Likewise, the Laboratoire d'Astrophysique de
Marseille (LAM) and the LETI (CEA) recently proposed to combine
active mirrors technology with flexible dies to generate deformable detectors \cite{ferrari}.
 This active mirror technology has achieved  variable curvature detectors, variable astigmatic detectors and
 also multi-mode detectors. The variable curvature mirror (VCM) technology consists of a circular membrane with a variable
thickness distribution. A force $F$, applied at the centre of the structure and perpendicularly to the membrane surface,
 results in an accurate spherical bending of the top surface
 of the membrane with a radius of curvature which depends
 on the applied force $F$. The thinned CMOS sensor attached on
 top of the VCM surface follows this curvature variation
  (convex to concave) \cite{chambion}. Another promising
  technology uses a tunable active mirror technology to bend
  the support to which the thinned CMOS sensor is attached
  (see \cite{laslandes}, developed for space missions).
This technology is able to generate many different shapes.
The prospects for using a curved detector for the MESSIER
pathfinder telescope are therefore more than promising.

\section{CONCLUSIONS}
In this paper we propose and analyse in detail a new
 optical design for a fast wide-field telescope based on
 the bi-folded reflective Schmidt scheme. This design provides high image quality over a large field of view, with minute distortion.
 The use of a flat folding mirror makes the design compact and eliminates spiders with their associated diffraction spikes and ghosts. The main factor which drives the optical design is the
 central obscuration. Applied to the optical scheme of
 the MESSIER pathfinder telescope, it has a $f$-ratio of 2.5 and provides nearly diffraction-limited image quality over
 an extended FoV of $1\fdeg6\times 2\fdeg6$, with compact PSF
 wings which are almost indistinguishable from the PSF
 of an unobstructed design.  With minor modifications the telescope is able to operate either with a curved detector or
 with a standard ordinary flat detector mounted in a cryostat and equipped with a set of standard photometric filters.
While the optical scheme consists of rather simple components,
the most complex part is the freeform primary mirror. However,
this mirror can be manufactured through existing active optics techniques and tested in a relatively simple setup using standard auxiliary optics.
This scheme will be used for the manufacturing of
a reduced prototype for the MESSIER telescope, and eventually
 for the full-scale instrument.

 This space mission
  will bring definitive answers to several
   key questions which have been recognised by various
   international panels (ESA's Cosmic Vision
   \cite{esa}, Europe's ASTRONET \cite{astronet}
   and USA's Decadal Survey \cite{decadal}) to be critical for our understanding of galaxy formation through the detection and characterisation of ultra-faint dwarf galaxies (which are predicted to be extremely abundant around normal galaxies, but which remain unobserved), and tracing the cosmic web (which feeds dark matter and baryons into galactic haloes) and which may contain the reservoir of missing baryons at low redshifts.

Funding Information.  ERC (European Research Council) (ERC-STG-2015--678777), LIA-ORIGINS (CNRS-CAS).

Acknowledgments. The authors acknowledge the support of the European Research council through the H2020-ERC-STG-2015--678777 ICARUS program, and the LIA-ORIGINS programme. DVG was supported as an Overseas Fellow of Churchill College, University of Cambridge, and an as International Senior Visiting Professor (Chinese Academy of Sciences). The authors thank Prof. Christopher Mihos for useful discussions on the
PSF of the Burrell Schmidt telescope.

%%%%%%%%%%%%%%%%%%%%%%%%%%%%%%%
% Full bibliography added automatically for Optics Letters submissions
% Note that this extra page will not count against page length
\ifthenelse{\equal{\journalref}{ol}}{%
\clearpage
\bibliographyfullrefs{sample}
}{}

%Manual citation list

\end{document}